\documentclass[12pt,manuscript]{aastex}

\usepackage{amsmath} 
\usepackage{natbib} 
\usepackage[normalem]{ulem} 
\usepackage{graphicx}
\usepackage{epstopdf}

\include{preamble}

\begin{document}

\def\bir{$XTE $ $J1810--197$ } 
\def\iki{$AX $ $J1845--0258$ } 
\def\uc{$SGR $ $1627--41$ }
\def\dort{$CXOU $ $J164710.2--455216$ } 
\def\bes{$1E $ $1547.0--5408$ } 
\def\alti{$SGR 0501+4516$ }

\newcommand{\rzero}{$r_{\mathrm{0}}$} 
\newcommand{\Smax}{$\Sigma$$_{\mathrm{max}}$ }
\newcommand{\Szero}{$\Sigma$$_{\mathrm{0}}$ } 
\newcommand{\dr}{$\Delta$r }

\def\la{\raise.5ex\hbox{$<$}\kern-.8em\lower 1mm\hbox{$\sim$}}
\def\ga{\raise.5ex\hbox{$>$}\kern-.8em\lower 1mm\hbox{$\sim$}} 
\def\be{\begin{equation}}
\def\ee{\end{equation}} 
\def\ba{\begin{eqnarray}} 
\def\ea{\end{eqnarray}} 
\def\be{\begin{equation}}
\def\ee{\end{equation}} 
\def\ba{\begin{eqnarray}} 
\def\ea{\end{eqnarray}} 
\def\d{$\partial$}
\def\R{$\right$} 
\def\L{$\left$} 
\def\a{$\alpha$} 
\def\Mdot*{$\dot{M}_*$}
\def\Mdotin{$\dot{M}_{\mathrm{in}}$} 
\def\Mdot{$\dot{M}$} 
\def\Pdot{$\dot{P}$} 
\def\Edot{$\dot{E}$}
\def\Omgdot{$\dot{\Omega}$} 
\def\Omgastdot{$\dot{\Omega}_{\ast }$} 
\def\Lin{$L_{\mathrm{in}}$}
\def\Rin{$R_{\mathrm{in}}$} 
\def\rin{$r_{\mathrm{in}}$} 
\def\rout{$r_{\mathrm{out}}$}
\def\rp{$r_{\mathrm{p}}$} 
\def\Rout{$R_{\mathrm{out}}$} 
\def\Omgast{$\Omega _{\ast }$}
\def\rA{$r_{\mathrm{A}}$} 
\def\Ldisk{$L_{\mathrm{disk}}$} 
\def\Lx{$L_{\mathrm{X}}$}
\def\Fx{$F_{\mathrm{X}}$}
\def\Lir{$L_{\mathrm{IR}}$} 
\def\Etot{$E_{\mathrm{tot}}$} 
\def\dE{$\delta E$} 
\def\dEb{$\DeltaE_{\mathrm{burst}}$} 
\def\dM{$\delta M$} 
\def\dEx{$\delta E_{\mathrm{x}}$}
\def\Bb{$\beta_{\mathrm{b}}$} 
\def\Be{$\beta_{\mathrm{e}}$} 
\def\Rc{\R_{\mathrm{c}}}
\def\dMin{$\delta M_{\mathrm{in}}$} 
\def\Teff{$T_{\mathrm{eff}}$}
\def\Tirr{$T_{\mathrm{irr}}$} 
\def\Tp{$T_{\mathrm{p}}$} 
\def\Firr{$F_{\mathrm{irr}}$}
\def\Tcrit{$T_{\mathrm{crit}}$} 
\def\rhot{$r_{\mathrm{h}}$} 
\def\rhmax{$r_{\mathrm{h,max}}$}
\def\Av{$A_{\mathrm{V}}$} 
\def\ah{$\alpha_{\mathrm{hot}}$} 
\def\ac{$\alpha_{\mathrm{cold}}$}
\def\p{$\propto$} \def\B*{$B_*$} 
\def\Bzero{$B_{\mathrm{0}}$}

%% /******************* %% ** The Header ** %%

\title{ON THE X-RAY OUTBURSTS OF TRANSIENT ANOMALOUS X-RAY PULSARS AND SOFT GAMMA-RAY REPEATERS}

\author{\c{S}IRIN \c{C}ALI\c{S}KAN\altaffilmark{1} AND {\"U}NAL ERTAN\altaffilmark{1} }
\altaffiltext{1}{Sabanc\i\ University, Orhanl\i - Tuzla, \.Istanbul, 34956, TURKEY}
%% /************** %% ** The Abstract ** %%

\begin{abstract}

We show that the X-ray outburst light curves of four transient anomalous X-ray pulsars (AXPs) and
soft gamma-ray repeaters (SGRs), namely $\bir$, $\alti$, $\uc$ and $\dort$, can be produced by the
fallback disk model that was also applied to the outburst light curves of persistent AXPs and SGRs
in our earlier work. The model solves the diffusion equation for the relaxation of a disk which has
been pushed back by a soft gamma-ray burst. The sets of main disk parameters used for these
transient sources are very similar to each other and to those employed in our earlier models of
persistent AXPs and SGRs. There is a characteristic difference between the X-ray outburst light
curves of transient and persistent sources. This can be explained by the differences in the disk
surface-density profiles of the transient and persistent sources in quiescence indicated by their
quiescent X-ray luminosities. Our results imply that a viscous disk instability operating at a
critical temperature in $\sim$ 1300 -- 2800 K range is a common property of all fallback disks around
AXPs and SGRs. The effect of the instability is more pronounced and starts earlier for the sources
with lower quiescent luminosities, which leads to the observable differences in the X-ray
enhancement light curves of transient and persistent sources. A single active disk model with the
same basic disk parameters can account for the enhancement phases of both transient and persistent
AXPs and SGRs. We also present a detailed parameter study to show the effects of disk parameters on
the evolution of the X-ray luminosity of AXPs and SGRs in the X-ray enhancement phases.
\end{abstract}


%% /************ %% ** Introduction ** %%

\section{Introduction}

Anomalous X-ray pulsars (AXPs) and soft gamma-ray repeaters (SGRs) are a special population of young
neutron stars whose rotational powers are not sufficient to account for their X-ray luminosities
($10^{34}-10^{36}$ erg s$^{-1}$, see Mereghetti 2008 for a recent review of AXPs and SGRs). The
spin periods of all known AXP and SGRs are in the range of 2 -- 12 s. These sources undergo short (
$<$ 1 s), super-Eddington soft gamma-ray bursts. Three out of four SGRs showed giant bursts with
energies greater than $10^{44}$ erg. After a soft gamma ray burst episode, (it is likely that some
of these bursts were missed), these sources enter an X-ray outburst/enhancement phase characterized by a sharp
increase and eventual decay in X-ray luminosity. Some of the AXPs and SGRs
have very low X-ray luminosities ($\sim 10^{33}$ erg s$^{-1}$) in the quiescent phase, and were
detected during these X-ray enhancement phases. These sources are called transient AXPs. During an
outburst, X-ray luminosity, \Lx, of the transient sources increases from a quiescent level of $\sim
10^{33}$ erg s$^{-1}$ to a maximum that remains in the \Lx ~range of persistent AXP/SGRs ($10^{34}$
-- $10^{36}$ erg s$^{-1}$).

Energetics and time scales of the soft gamma-ray bursts which are very likely to have magnetic origin
resulted in the classification of such objects as ``magnetars'' (Duncan \& Thompson 1992). In the
magnetar model, the source of the X-ray luminosity is the magnetic field decay, and the rotation
rate of the neutron stars in these systems is assumed to be slowing down by the magnetic dipole
torques in vacuum. This requires that the dipole component of the magnetic field has magnetar
strength (\Bzero~$> 10^{14}$ G) on the surface of the neutron star. The magnetar model has no
explanation for the period clustering of AXP/SGRs. Explaining the optical and infrared (IR)
observations of persistent and transient AXP/SGRs in quiescent and outburst (enhancement) phases
within the magnetar model also poses problems.

The fallback disk model (Chatterjee et al. 2000; Alpar 2001) was initially proposed to explain the
spin periods and X-ray luminosities of AXPs and SGRs. It was suggested that the initial
properties of fallback disks, together with magnetic dipole moment and initial spin period, could be
responsible for the formation of the other young neutron star populations as well (Alpar 2001).
Later, it was shown that the optical, IR and X-ray observations of persistent AXPs and SGRs
(hereafter, we use "AXPs" to denote both AXPs and SGRs) in both quiescent and enhancement phases can
be explained consistently by the presence of active, accreting fallback disks in these systems
(Ek\c{s}i \& Alpar 2003; Ertan \& Alpar 2003; Ertan \& Cheng 2004; Ertan et al. 2006; Ertan \& \c{C}al{\i}\c{s}kan 2006).
The detection of AXP 4U 0142+61 in mid-IR bands clearly indicates the presence of a disk around this
source (Wang et al. 2006). This mid-IR data, together with earlier detections in optical and near IR
bands, can be well fit by an irradiated active disk model, provided that the dipole field which
interacts with the accretion disk has conventional values of \Bzero~ $\simeq$ $10^{12}-10^{13}$ G
on the surface of the young neutron star (Ertan et al. 2007). Coherently with these results, X-ray
luminosity, period, period derivative, and statistical distribution of AXPs can also be produced by
the evolution of the neutron stars with fallback disks and with dipole fields \Bzero~$< 10^{13}$ G
(Ertan et al. 2009). Based on these constraints on \Bzero~indicated by our results, we proposed that the
strong magnetic fields of AXPs must thus reside in multipoles which die rapidly in strength with
increasing distance from the neutron star (Ek\c{s}i \& Alpar 2003;
Ertan at al. 2007, 2009). Recently reported
upper bound on the period derivative of SGR 0418+5729 unambiguously revealed that the dipole field
strength of this source cannot be greater than $\sim$ 7 $\times 10^{12}$ G on the surface of the
neutron star (Rea et al. 2010). This is in full agreement with our explanation, and clearly shows
that the soft gamma-ray bursts do not require magnetar dipole fields. Furthermore, if the dipole 
field is below this upper limit, then the dipole spin down age would not be accurate and other torque and 
magnetic field effects would need to be taken into account. In the frame of the disk
model, rotational properties and X-ray luminosity of this SGR can be reached simultaneously within
the cooling timescale of a neutron star with \Bzero~$\simeq$ 10$^{12}$ G (Alpar et al. 2011).
Recently, Tr\"{u}mper et al. (2010) showed that the high-energy spectrum of AXP 4U 0142+61 
can be produced in the accretion column of this source mainly by the bulk Comptonization process.

In the present work, we investigate the X-ray enhancement (outburst) light curves of persistent and
transient AXPs. We pursue the results of the work by Ertan \& Erkut (2008) on the X-ray
outburst light curve and the spin evolution of the transient AXP XTE J1810--197. The X-ray outburst
light curve of this source showed a different decay morphology than those of persistent sources
(Ibrahim et al. 2004, Bernardini et al. 2009). By means of model fits to the X-ray enhancement data,
Ertan \& Erkut (2008) concluded that this difference could be due to a viscous disk instability (see
e.g. Lasota 2001 for a review of the disk instability model (DIM)) in the fallback disks at a critical
temperature in the $\sim 1000-2000$ K range. The fallback disks around AXPs are expected to have
similar chemical compositions. If one of the AXP disks undergoes a thermal-viscous disk
instability at a particular critical temperature, then the others are also expected to show the same
instability at the same temperature. Our aim is to test this idea by applying the same model to the
X-ray outburst data of other transient AXPs.

There are some difficulties in testing our model when the observed X-ray luminosity is close to the
quiescent level (\Lx~$\sim 10^{33}$ erg s$^{-1}$) of the transient AXPs. As \Lx~decreases,
temperature also decreases and thus effects of interstellar absorption increase. Furthermore, at these low
temperatures, a significant fraction of the X-ray emission may come from outside the observational
X-ray band and estimates of the bolometric luminosities are model dependent (Gotthelf \& Halpern
2007; Bernardini et al. 2009). Our model gives the total accretion luminosity without addressing the
X-ray spectrum from the surface or the accretion column of the neutron star. For comparison with
data, we assume that the observed \Lx~ is a close representation of the total \Lx, which is a good
assumption for the X-ray luminosities down to a few times $10^{33}$ erg s$^{-1}$ but does not allow
a reliable comparison for lower luminosities. Another difficulty at very low \Lx~arises due to the
fact that the luminosity contribution from the intrinsic cooling of the neutron star (Page et al.
2006) could become comparable to the accretion luminosity depending on the age of the source.
Keeping these uncertainties in mind, we extend the model curves to the quiescent luminosity levels
to present the model predictions at low \Lx. As in other works on disk accretion, we do not perform
$\chi^2$ tests, since it is misleading due to the uncertainties in the disk, like the local instabilities 
close to the inner disk, that are not possible to address in the models.

Basic disk parameters, namely, the critical disk temperature, kinematic viscosity parameters,
irradiation strength, and the radius dependence of the surface density of the extended disk are
expected to be similar in the fallback disks of different AXPs. This forces us to a difficult task
of producing the X-ray outburst light curves of AXPs with a single set of these basic parameters. We
describe our model and discuss the effect of the disk parameters on the X-ray luminosity
evolution in Section 2. Properties of the transient AXPs that were observed in X-ray enhancement phases
are summarized in Section 3. We discuss the results of the model calculations in Section 4 and summarize our
conclusions in Section 5.

\section{The Numerical Model}

\subsection{Description of the Model Parameters}

We solve the disk diffusion equation (see, e.g., Frank et al. 2002) in the way described in Ertan \&
Alpar (2003) and Ertan \& Erkut (2008). In this model, it is assumed that a soft gamma-ray burst
triggered on the surface of the neutron star pushes the inner disk matter outward. Some of this matter 
could escape from the system, while the remaining part creates a surface
density gradient at the innermost disk. The resultant pile-up, centered at \rzero, and the
underlying disk distribution are represented by a Gaussian $ \Sigma = \Sigma_{\mathrm{max}}$ exp
[$-(r-r_0)^{2}/\Delta r^{2}] $ and a power-law $\Sigma = \Sigma_{\mathrm{0}}~(r_{\mathrm{in}}/r)^p$
surface density profile respectively as the initial condition of the model.

We assume that the surface density profile of the 
innermost disk in the quiescent state is close 
to  the standard thin disk profile $\Sigma \propto r^{-3/4}$, 
and the disk matter from \rin~to a radius $r_1$, with mass 
$\delta M$, 
is initially pushed out to a narrow radial region at radius 
$r_1$ by the burst. 
As the matter with a range of specific angular momentum mixes 
and piles up at $r_1$, the angular momentum is redistributed
rapidly due to mixing and the narrow radial extent of the pile-up. 
The required timescale for the sharing of angular momentum is 
of the order of a few $h/v_K$, where $v_K$ is the 
Keplerian velocity close to $r_1$ and $h$ is the thickness of 
the disk, typical lengthscale for efficient viscous interaction, 
at $r_1$. 
Taking $h(r_1) \sim 10^{-2}r_1$ and $r_1 \sim 10^{10}$ cm, $h/v_K \simeq 10^{-2}
\Omega_K^{-1}$ is found to be a fraction of a second.
This implies that the angular momentum is effectively shared 
during the formation of the pile-up.
After the 
burst episode, this matter circularizes  
at a radius between \rin~and $r_1$ depending on the mean specific angular 
momentum of the pile-up.  The matter spreading from 
this circularization radius to both inner and 
outer radii with a surface density peak close to the circularization 
radius is represented by a Gaussian 
as the initial mass distribution in our model. 
The center \rzero~of the Gaussian in the model 
could be assumed to represent this circularization radius. 

Our model light curves do not 
sensitively depend on \rzero~or the details of the Gaussian 
distribution. Similar model light curves could 
be produced with different \rzero~values (within a factor 
of a few), provided that $\delta M$ contained in 
the Gaussian distribution remains the same. This density 
gradient leads to an abrupt rise in the mass-flow 
rate at the innermost disk. The rise phase of the X-ray 
light curve is produced by the enhanced mass-flow 
rate to the Alfv\'en radius and subsequently onto the surface 
of the neutron star. 
Since the exact position of the inner disk radius, \rin, 
does not change the rate of mass inflow, for simplicity 
we take \rin~constant and equal to the inner disk radius 
in quiescence. The mass accretion rate in the decay phase of the light curve is 
governed by the viscous relaxation of the inner disk matter.
At the end of the decay phase, the luminosity 
converges to the quiescent X-ray 
luminosity level of the source. The X-ray luminosity produced 
by the inner disk through viscous dissipation is 
negligible compared to the luminosity powered by mass accretion onto the surface of 
the neutron star.

The evolution of the disk is determined by solving the diffusion 
equation (see, e.g. Frank et al. 2002).  While \dM~is
important in the evolution of the X-ray luminosity, the detailed shape of the Gaussian does not
significantly affect the model light curve in the long term (Ertan et al. 2006). For a viscously
evolving disk, the power index $p$ of the extended surface density
profile is expected to be $\sim 3/4$ (Frank et al. 2002); we take $p$ = 3/4 in our calculations. 

We keep the inner radius of the disk \rin~constant at a value near the Alfv\'en radius for a dipole
magnetic field with strength \Bzero~= 10$^{12}$ G on the surface of the neutron star. Since the
model fits are not sensitive to \rin, we are not able to constrain \Bzero. Our results are not
sensitive to the value of the outer disk radius \rout~either, since the viscous timescale along the disk is
much longer than the enhancement episodes of AXPs. In our calculations we take \rout~= 10$^{13}$
cm.

Irradiation parameter $C$ represents the efficiency of X-ray irradiation flux $ F_{\mathrm{irr}} =
(C \dot{M} c^{2})/ (4 \pi r^{2}) $ (Shakura \& Sunyaev 1973) where \Mdot~is the mass accretion rate
onto the neutron star. Infrared and optical data of the persistent AXPs can be accounted for by an
active disk model with $C$ in the range 10$^{-4}$ -- 10$^{-3}$ (Ertan \& \c{C}al{\i}\c{s}kan 2006)
similar to those estimated for low-mass X-ray binaries. Total X-ray luminosity is related to the
mass accretion rate through $ L = G M \dot{M}/r. $ We take $f$ = \Mdot / \Mdotin~= 1, where \Mdotin~is the
mass-flow rate arriving at the innermost radius of the disk. Actually, $f$ could be less than unity, that is, 
a fraction of \Mdotin~can escape the system or may not return back to the disk. Employing lower $f$ 
values in the calculations does not change our qualitative results, but requires a modification of some 
other disk parameters. This will be discussed in Section 4. For comparison of the model with
observations, we take the X-ray luminosities in the observational bands to represent the total X-ray
luminosity of the source. 

Different viscosity states prevail in the hot ($T >$ \Tcrit) and cold ($T <$ \Tcrit) regions of the
disk. For the kinematic viscosity, we use the \a-prescription (Shakura \& Sunyaev 1973) with \a~=~\ac~ and
\a~=~\ah~ in the cold and hot viscosity states respectively. For a review of the DIMs, see e.g. Lasota 2001. The disk evolution model we use here is the same as DIMs of
LMXBs and dwarf novae (DNs). The difference is the value of the critical temperature, \Tcrit. In LMXB and DN disks,
the thermal-viscous instability operates around the ionization temperature of hydrogen
(\Tcrit~$\sim$ 10$^{4}$ K). In the case of AXPs, both the temperature profile and the chemical
composition of the disk are different from the hydrogen disks of LMXB and DNs. In AXP disks, metallicity is
likely to be much higher than in an LMXB disk. We should also note that the hottest,
innermost parts of the LMXB disks do not exist in AXP disks due to stronger dipole fields of AXPs
that cut the disk at a relatively larger radius ($\sim$10$^{9}$cm).

The critical temperatures depend sensitively on the details of the ionization properties of the disk
matter. Independent of these details, if the disk undergoes a global disk instability at a
particular temperature, the resultant evolution of the disk produces a light curve which can be
easily distinguished from the pure viscous decay curve (not affected by instabilities). Furthermore,
fallback disks around AXPs are likely to have similar chemical compositions and similar critical
temperatures.

In our model, there are five main disk parameters, namely, \Tcrit, \ac, \ah, $p$, and
$C$, which govern the evolution of the accretion disk for a given initial mass distribution.
Among these, \Tcrit~and $C$ are degenerate parameters. There is a constraint on the range of 
$C$ obtained in our
earlier work on the persistent AXPs (Ertan \& \c{C}al{\i}\c{s}kan 2006). These basic disk parameters
are very likely to have similar values for fallback disks of different AXPs. In Section 2.2, we
investigate the effects of model parameters on the evolution of the disk to clarify the subsequent
discussion on the light curves of persistent and transient sources.

\subsection{Parameter Study} 

Observations of X-ray outburst (enhancement) light curves of different AXPs with different
energetics and time-scales provide an opportunity for a detailed test of the fallback disk model 
and also for constraining the model parameters. In this section, we investigate the
effect of important disk parameters on the X-ray outburst light curves of model sources.

\subsubsection{ Different Burst Energies } Soft gamma-ray bursts of AXPs are likely to have magnetic
origin and to occur close to the neutron star. Assuming an isotropic emission, a small fraction $
\delta{E} / E_{\mathrm{tot}} \sim H_{\mathrm{in}}/r_{\mathrm{in}} $ of the total burst energy,
\Etot, is absorbed by the disk where \dE~ is the part of the burst energy illuminating the disk and
$H_{\mathrm{in}}$ is the half-thickness of the disk at $r=r_{\mathrm{in}}$. $H/r$ is roughly
constant along the disk and is about 10$^{-3}$ in the accretion regime of AXPs. When the inner disk
is pushed back and heated by \dE, part of the inner disk matter could escape the system, while the
remaining part piles up, forming a surface density gradient at the inner disk (see Ertan et al. 2006
for details). In our model, the Gaussian surface density distribution represents the inner pile-up
and the power-law surface density distribution stands for the outer extended disk that is expected
to remain unaffected by the soft gamma-ray burst. The position \rzero~and the total mass \dM~of the
pile-up, for a given \dE, depend on the inner disk radius \rin~ and the mass distribution of the
inner disk just before the burst event. In steady state, the surface density profile of the inner
disk-magnetosphere boundary is not well known. Assuming that the inner disk conditions are similar
for fallback disks of AXPs, it is expected that a higher burst energy pushes the inner disk to a
larger radius, and creates a greater density gradient at the inner disk.

In the quiescent state, the mass-flow rate, \Mdot, decays very slowly and therefore can be taken as
constant in the models. In this steady state, $ \dot{M} \propto \Sigma~ \nu $ where $ \nu = \alpha~
c_{S}~ H $ is a function of temperature and radius, and depends also on the ionization properties of
the disk matter. The pressure-scale height of the disk $ H \simeq c_{S}/\Omega_{K}, $ then $ \nu
\propto T~r^{3/2} \propto r^{3/4} $ (see, e.g. Frank et al. 2002) where $T$ is the mid-plane
temperature of the disk. The irradiation temperature  \Tirr~$\propto r^{-1/2} $ modifies the
effective temperatures and the stability criteria of the disk without significantly affecting the
mid-plane temperatures in the accretion regime of AXPs and SGRs (e.g. Dubus et al. 1999).  Then, in 
the quiescent state, the surface density
of the disk  $\Sigma$ $\propto$ $r^{-p}$ with $p$ = 3/4.
The main role of the irradiation is to slow down the decay of the X-ray luminosity, preventing the 
rapid propagation of the cooling front inward. This will be investigated in detail in Section 2.2.3.

We first illustrate X-ray enhancement light curves of a model source with different \dM~values
representing the evolution of the same source with different burst energies. In the first exercise,
we compare the model curves without invoking the instability (pure viscous evolution) with $\alpha$
= 0.1, \rin = $10^{9}$ cm, and \rout = $10^{13}$ cm. Three different illustrative light curves
presented in Figure 1 are obtained with different \dM~values that give peak luminosities of
1 $\times$ 10$^{36}$, 3 $\times$ 10$^{36}$, and 1 $\times$ 10$^{37}$ erg s$^{-1}$. The quiescent
\Lx~of all these sources are close to 10$^{35}$ erg s$^{-1}$, a typical luminosity of a
persistent AXP. It is seen that the X-ray luminosities follow almost the same decay curve after
$\sim$ a few months and eventually reach their quiescent level. For the first several weeks of the
outburst both the fluences and the functional forms of the decay curves are very different from each
other. For this initial decay phase, the model curves can be fit by a power law $ L =
L_{\mathrm{peak}}~(t/t_{\mathrm{peak}})^{-n} $ with power-law indices of 0.88, 0.68 and 0.46. The minimum burst energy
imparted to the disk can be estimated from \dM~using $ \delta{E} \simeq G~ M~ \delta{M}~
(r_{\mathrm{in}}^{-1} - r_{0}^{-1}) \simeq G~ M~ \delta{M}~ /~ r_{\mathrm{in}}. $ For these illustrative models, \dM~=
4.9 $\times$ $10^{22}$, 2.1 $\times$ $10^{22}$, and 8.9 $\times$ $10^{21}$ g, and the estimated
\dE~values are 9.2 $\times$ $10^{39}$, 3.8 $\times$ $10^{39}$, and 1.7 $\times$ $10^{39}$ erg,
respectively. Note that actually a higher \dE~accumulates a greater
\dM~at a larger \rzero. Since the chosen \rzero~does not significantly affect the light curve, for
simplicity, we take \rzero~constant (5 $\times$ $10^{9}$ cm) for all these illustrative simulations.

We repeat the same calculations for a model source with a quiescent luminosity around 10$^{33}$ erg
s$^{-1}$, typical for transient AXPs. In Figure 2, we present three different light curves produced
by pure viscous evolution of the disk for three different \dM~ values, without changing the other
parameters. In this case, estimated \dE~values are 4.5 $\times$ $10^{38}$ erg, 2.3 $\times$
$10^{38}$ erg and 1.2 $\times$ $10^{38}$ erg respectively. For these sources, first $\sim$ 100 days
of the decay curves can be fitted by power laws with indices 0.91, 0.73, and 0.59. In Figure 2, like the model
sources given in Figure 1, the sources with higher \dM~show higher peak luminosities and sharper
decay curves in this early phase of evolution. 

In these examples, the model light curves 
are produced by pure viscous evolution of the disk without any instability. Comparing Figures 1 and 2, we
might conclude that the sources with similar \Lx~in quiescence could show decay curves with rather
different power-law indices, while it is also possible that sources with different quiescent
luminosities could give similar decay curves in the early decay phase of the outburst. In Section 2.2.2, 
we show that this early phase of the X-ray light curves can indeed be produced by pure viscous 
evolution of the disk for both transient and persistent sources. This pure viscous evolution model 
gives similar long-term curves for all model sources (Figures 1 and 2), which
would be the case if there were no critical temperature leading to viscous instability in the disk.
In the following sections, we show how the presence of a critical temperature leads to systematic
differences in the functional form of the decay curves depending on the X-ray luminosities in
quiescence.

\subsubsection{Quiescent X-Ray Luminosity and Critical Temperature}

The main characteristics of X-ray outbursts of soft X-ray transients (SXTs) and DNs
can be successfully accounted for by DIMs (see, e.g., Lasota 2001 for a
review). In these systems, the viscous instability manifests itself at temperatures around 10$^{4}$
K which corresponds to the ionization temperature of hydrogen. In the DIMs, disk regions with
temperatures higher and colder than this critical temperature are in hot and cold viscosity states
respectively. Different $\alpha$ parameters are employed in the cold and hot states (\ah~$\sim$ 0.1
and \ac~$\sim$ 0.01 -- 0.05, see Section 2.2.4) to obtain reasonable model fits to the X-ray outburst
light curves of SXTs and DNs. We note that these viscosities are turbulent in both hot and cold
states.

We now investigate the effect of viscous instability with different critical temperatures on the
model X-ray light curves of two illustrative sources with quiescent X-ray luminosities of 10$^{33}$
and 10$^{35}$ erg s$^{-1}$ as representatives of transient and persistent AXPs. For both model
sources, we take $C$ = 1 $\times$ 10$^{-4}$. The results are seen in Figure 3. Panel (a)
shows the luminosity evolution of a persistent source, for three different \Tcrit~values, as well
as the pure viscous decay (no viscous instability). Similarly, panel (b) shows the
evolution of a transient source. For a given source, the model curves with higher \Tcrit~values
diverge from pure viscous decay curve earlier. Comparing Figures 3(a) and (b), we see that for a
particular \Tcrit, the light curve of the persistent sources (high quiescent luminosity) deviate
from the pure viscous decay curve much later than the transient sources (low quiescent luminosity).
For instance, for \Tcrit~= 2000 K, the light curve of the transient source deflects from the
pure viscous decay curve at $t$~$\sim$~100 days. 
For the persistent source with the same \Tcrit, the deviation starts at
$t$~$\sim$~200 d and the luminosity decreases much slower compared to the transient source.
For \Tcrit~$\sim$~1500 K, the light curve of the persistent source is
indistinguishable from the pure viscous decay until $t$~$\sim$~400 days, while for the transient source,
the deviation starts as early as $t$~$\sim$~200 days (Figure 3). This characteristic difference in the light curve
morphologies of high and low-luminosity AXPs in the decay phase is mainly due to the differences in the surface density
and temperature profile of the disks in the quiescent states. These properties could be estimated
from the X-ray luminosities in quiescence, which scales with the accretion rate onto the neutron
star.

\subsubsection{X-Ray Irradiation}

Another factor that plays an important role in the evolution of the disk and the X-ray luminosity is the
X-ray irradiation of the disk. The irradiation flux can be written as $ F_{\mathrm{irr}} = C \dot{M}
c^{2}/ 4 \pi r^{2} $ where irradiation efficiency $C$ depends on the albedo of the disk faces and
the irradiation geometry (Shakura \& Sunyaev 1973). The critical temperature discussed in Section 2.2.2
and $C$ are degenerate parameters. The results of our earlier work on the X-ray and IR data of AXPs
constrain the value of $C$ to the range 10$^{-4}$ -- 7 $\times$ 10$^{-4}$ (Ertan \&
\c{C}al{\i}\c{s}kan 2006), which remains in the range of the estimated irradiation efficiencies of
SXTs (10$^{-4}$ -- 10$^{-3}$; de Jong et al. 1996; Dubus et al. 1999; Tuchman et al. 1990).

The effective temperature of a steady disk at a radial position is determined by the dissipation
rate given by $ D = 9~ \nu~ \Sigma~ \Omega_{K}^{2}~ /~ 8, $ where $\Omega_{K}$ is local Kepler
velocity, and by the X-ray irradiation flux \Firr. Including \Firr, the effective temperature can be
written as $ \sigma T_{\mathrm{eff}}^{4} \simeq D + F_{\mathrm{irr}}. $ 
Since $\nu \Sigma \propto \dot{M}$,  both \Firr~and $D$ have the same \Mdot~(and thus \Lx) dependence. 
The irradiation flux and the dissipation rate decrease with radial distance as $r^{-2}$ and $r^{-3}$ 
respectively. Equating \Firr~to $D$, we find $r \sim 2 \times 10^{9}$ cm which does not depend on \Mdot. 
For smaller radii, $D$ dominates \Firr, while \Firr~is the dominant source of heating beyond this radius
(see, e.g. Frank et al. 2002).

The radius \rhot~of the hot inner disk is determined by the strength of the X-ray
irradiation ($T$($r$=\rhot) = \Tcrit). Increasing (decreasing) \Firr~increases (decreases) the
effective temperatures at all radii, and \rhot~is situated further out (in). This implies that the
sources with higher X-ray luminosity have greater \rhot. This is actually the main reason that leads
to different X-ray light curve morphologies in the enhancement phases of transient and persistent
sources. The rate of mass accretion which powers \Lx~is determined by the surface density profile of
the disk. The sources with higher surface densities have higher accretion rates, higher X-ray
irradiation fluxes, and thus greater \rhot. To illustrate, for \Tcrit~=~1500 K and $C =$ 1.5
$\times$ 10$^{-4}$, we find \rhot~= 1.4 $\times$ 10$^{11}$ cm for \Lx~= 10$^{35}$ erg s$^{-1}$ and
\rhot~= 1.4 $\times$ 10$^{10}$ cm for \Lx~= 10$^{33}$ erg s$^{-1}$.

At the beginning of the X-ray enhancement phase, the innermost disk that was emptied by the burst is
refilled rapidly due to high density gradients, leading to a sharp rise in X-ray luminosity. The
mass-flow rate in the inner disk, the accretion rate onto the neutron star, and thereby the X-ray
luminosity, \Lx, and \rhot~reach their maximum values. Subsequently, \rhot~decreases gradually at a
rate governed by the decreasing X-ray flux.

In the quiescent phase, the mass-flow rate at the inner hot disk depends on the conditions at the
cold outer disk. The hot disk easily transfers all the mass flowing from the outer cold disk toward
\rin. During an enhancement, the mass-flow rate is determined mainly by the viscous processes at the
hot inner disk. Just before the onset of the X-ray outburst, the total amount of mass that remains
within \rhot, the position of \rhot~ and the rate at which it moves inward all affect, and also
depend on the evolution of \Lx.

Initially, the light curve mimics that of a pure viscous decay, since the information from
\rhot~moving inward reaches the inner disk after a viscous time-scale across the hot disk. With
decreasing \rhot, the total hot mass contributing to the accretion with high viscosity also
decreases. After the conditions at \rhot~ start to modify \Mdotin, \Lx~decreases more rapidly and
diverges from the pure viscous decay curve.

In Figure 4, we give the model curves with different $C$ values, keeping \Tcrit~= 1750 K for all the
simulations. Comparing with Figure 3, it is seen that the light curves for different $C$ values are
similar to those obtained with different \Tcrit~values, keeping $C$ constant. It is also seen that
the viscous instabilities triggered at the same \Tcrit~and with the same $C$ produce very different
light curves for transient and persistent sources.

\subsubsection{Viscosity Parameter} 

For both transient and persistent AXPs, the rise, turnover and early decay phase (several weeks to
months) of the X-ray light curve are produced by the evolution of the hot inner disk matter and the 
resultant accretion onto the neutron star. In all
our calculations, we take \ah~= 0.1 as in our earlier works. For all the sources, this initial phase
of the light curve is indistinguishable from that produced by a pure viscous evolution of the disk,
that is, the evolution of a disk remaining in the same viscosity state at all radii. By illustrative
model light curves (Figures 3 -- 5), we have shown that the deviation from this pure viscous decay
phase starts much earlier in transient AXPs which have relatively low luminosities in the quiescent
phase. After the instability starts to affect the accretion rate, the value of \ac~has an important 
role in the
evolution of the X-ray luminosity \Lx. From the DIMs of SXTs and DNs, \ac~is estimated to be $\sim$
0.01 -- 0.05 (Lasota 2001). From model fits to X-ray enhancement light curve of $\bir$, Ertan \&
Erkut (2008) found that \ac~$\sim$ 0.03 with \Tcrit~$\sim$ 1500 K produce reasonable model curves.
In the present work, we also refine the model parameters of Ertan \& Erkut (2008) through a
comparative study with the X-ray enhancement light curves of other transient AXPs, including the new
data points of $\bir$ (Section 3).

Keeping all the other parameters constant, we see that small variations in \ac~could lead to
significant changes in the light curve at the end of the decay phase. To illustrate this effect, we
present model curves with different \ac~values in Figure 5.
The depths of the minima at the end of the model light curves of the model sources depend mainly on the
values of \ac. It is seen in Figure 5 that the model light curves settle
down to the quiescent level following quite different morphologies even for small changes in \ac.

The physical reason producing the minima in the model curves can be summarized as follows: 
after the formation of the
pile-up at the inner disk, the accretion rate abruptly increases in the hot disk region ($r <$
\rhot) due to newly formed density gradients. The resultant increase in \Lx~pushes \rhot~to larger
radii, causing part of the previously cold disk region to enter the hot viscosity state. Due to the
density gradients and more efficient kinematic viscosity in the hot state, the inner disk matter is
depleted at a rate much higher than the mass-flow rate provided by the outer disk. As a result,
surface density profile $\Sigma$($r$) of the inner disk decreases below the extrapolation of
$\Sigma(r$) of the cold outer disk. Meanwhile, \Lx~decreases due to both decreasing $\Sigma$ of the
inner disk and the propagation of \rhot~inward with decreasing \Lx.

The rate of refilling of the innermost disk regions sensitively depends on the value of \ac. It is
seen in Figure 5 (bottom panel) that the minimum in the model light curves becomes more pronounced for smaller
\ac~values. This is because the surface density gradients are smoothed out more rapidly with higher
kinematic viscosity. Observed X-ray enhancement light curves provide an opportunity to constrain the
value of \ac. Nevertheless, the limitation for testing the models can also be clearly seen in Figure
5 (bottom panel). The accretion luminosity of the transient AXPs might decrease even below the intrinsic cooling
luminosity of the neutron star depending on the age of the source. 

\subsubsection{Outer Disk Radius}

The outer disk radius, \rout, defines the extent of the active accretion disk and depends on the
minimum disk temperature at which the disk becomes inactive. X-ray luminosity, rotational and
statistical properties of AXPs can be explained by the long-term evolution of neutron stars evolving
with fallback disks that become inactive at low temperatures around $\sim$100--200 K (Ertan et al.
2009). In this model, \rout~gradually decreases in time with slowly decreasing quiescent X-ray
luminosity of the source.

The evolution of \rout~has an important effect on the long-term (10$^{3}$ -- 10$^{5}$ yr) evolution of
AXPs. Nevertheless, the position of \rout~does not affect the X-ray enhancement light curves of
AXPs, which last from months to several years, . The outer radii of the fallback disks of known AXPs
are estimated to be greater than about a few $\times$ 10$^{12}$ cm (Ertan et al. 2007). Viscous timescale across
the disk is longer than the duration of the enhancement phase. In our calculations, we set \rout~=
10$^{13}$ cm. We note that the radius \rhot~of the hot disk which is the border between low and high
viscosity regions of the disk should not be confused with \rout.

\section{Application of the Model to the X-ray Enhancement Light Curves of Transient AXPs} 

Our model parameters and their effects on the evolution of the sources are described in detail in
Section 2. Now, we test this model, performing model fits to the X-ray outburst light curves of the
sources $\bir$, $\uc$, $\dort$ and $\alti$ (Figures 6 -- 9). The model parameters are presented in
Table 1. All these sources were detected in the decay phases of their X-ray outbursts. The rise and
turn-over phases of the outburst were missed.

The X-ray flux of $\bir$ in quiescence (during 1993 -- 1999) was 5.5 $\times$ 10$^{-13}$ erg
cm$^{-2}$ s$^{-1}$ (Gotthelf et al. 2004) and increased to about 5.5 $\times$ 10$^{-11}$ erg
cm$^{-2}$ s$^{-1}$ during the outburst (Ibrahim et al. 2004). Most recent distance and the
corresponding peak luminosity estimates for this source are $d$ = 5 kpc, \Lx~= 1.3 $\times$ 10$^{36}$
erg s$^{-1}$ (Ibrahim et al. 2004), $d$ = 3.3 kpc, \Lx~= 5.8 $\times$ 10$^{35}$ erg s$^{-1}$
(Lazaridis et al. 2008) and $d$ = 3.5 kpc, \Lx~= 6.6 $\times$ 10$^{35}$ erg s$^{-1}$ (Bernardini et
al. 2009). In our calculations, we take $d$ = 3.5 kpc. For the model fits, we use 2 -- 10 
keV XTE data (Ibrahim et al. 2004) and 0.6 -- 10 keV $XMM$ data (Bernardini et al. 2009). The $XMM$ 
data were converted to 0.1 -- 10 keV unabsorbed flux with \emph{WebPIMMS}, using the 3BB model 
described in their paper. The XTE data, given in counts s$^{-1}$ PCU$^{-1}$, were converted to unabsorbed flux 
using a conversion factor. The factor was chosen so as to align the first $XMM$ data with the 
corresponding XTE data (in 2003 September). Our model curve is given in Figure 6. For all the 
sources, together with the best model fits, we also plot the pure viscous decay curves for 
comparison.

The transient $\uc$ underwent an X-ray outburst in 1998 and its decay curve is also similar to those
of other transient sources (Mereghetti et al. 2006). The peak luminosity of $\uc$ was
9.5 $\times$ 10$^{34}$ d$_{11}^{2}$ erg s$^{-1}$ during the outburst and the distance was measured
as 11.0 $\pm$ 0.3 kpc (Corbel et al. 1999). The source subsequently decayed to quiescence with
\Lx~$\sim$ 3.9 $\times$ 10$^{33}$ d$_{11}^{2}$ erg s$^{-1}$ (Kouveliotou et al. 2003). In 2008 May,
a new X-ray outburst was observed in $\uc$ (Palmer et al. 2008). The absorbed 2--10 keV flux was
$\sim$1.3 $\times$ 10$^{-12}$ erg cm$^{-2}$ s$^{-1}$, corresponding to a luminosity of
\Lx~$\sim$3 $\times$ 10$^{34}$ d$_{11}^{2}$ erg s$^{-1}$ (Esposito et al. 2009). There is 
an uncertainty in estimating the unabsorbed flux of this source due to high interstellar absorption 
with $N_{\mathrm{H}} \sim 10^{23}$ cm$^{-2}$ (Mereghetti et al. 2009).
We take $d$ = 11 kpc to estimate the luminosity. Our model curve is seen in Figure 7.

A soft gamma-ray burst from $\dort$ was detected with $Swift$ BAT on
2006 September 21 (Krimm et al. 2006). It was observed for $\sim$~20 ms with total energy $\sim$ 
3 $\times$ 10$^{37}$ erg (15 -- 150 keV, for $d$ = 5 kpc; Muno et al. 2007).
The observed maximum X-ray flux data point was reported 1.6 days after the burst.
The X-ray flux data of AXP $\dort$ cover about 150 days on the decay phase of its outburst in
2006 September (Israel et al. 2007, Woods et al. 2011). The X-ray luminosity of the source increased
from $\sim$ 1 $\times$ 10$^{33}$ d$_{5}^{2}$ erg s$^{-1}$ to more than 1 $\times$ 10$^{35}$
d$_{5}^{2}$ erg s$^{-1}$ during the outburst (Muno et al. 2007). The distance of $\dort$, located in
a star cluster, was estimated as 2 kpc $<$ $d$ $\le$ 5.5 kpc (Clark et al. 2005). We convert the 
flux data to luminosity using $d$ = 5 kpc.
The data seen in Figure 8 seem to be taken in the early decay phase of this source, and therefore does not 
constrain \Tcrit~yet.

Another transient source that was discovered in an outburst is $\alti$ (Barthelmy et al. 2008). 
Subsequently, the source
was observed with $XMM-Newton$, $Swift$ and $Suzaku$ in the decay phase, starting from a maximum
absorbed 1 -- 10 keV flux
of 4.1 $\times$ 10$^{-11}$ erg cm$^{-2}$ s$^{-1}$ (Rea et al. 2009; Figure 9). The distance of
$\alti$ is not very well determined, with estimates $d$ = 1.5 kpc (Aptekar et al. 2009), $d$ = 4 kpc 
(Nakagawa et al. 2011), $d$ = 5 kpc (Rea et al. 2009) and $d$ = 10 kpc (Enoto et al. 2009; Kumar et al.
 2010). The pre-outburst quiescent (0.1 -- 2.4 keV) flux of this source
was reported as $\sim$ 4.1 $\times$ 10$^{-12}$ erg cm$^{-2}$ s$^{-1}$ (Rea et al. 2009), which
corresponds to a 1 -- 10 keV flux of 1.3 $\times$ 10$^{-12}$ erg cm$^{-2}$ s$^{-1}$. Assuming a 
distance of 5 kpc, the maximum and quiescent luminosities are 1.2 $\times$
10$^{35}$ erg s$^{-1}$ and 4 $\times$ 10$^{33}$ erg s$^{-1}$ respectively. We take $d$ = 5 kpc in our calculations.

The transient AXP $\bes$ at the end of the $\sim$ 100 days of decay, showed re-brightening (Camilo et
al. 2008), possibly due to another soft gamma-ray burst, which we could not address in the model. 
Long-term behavior of this source can be
studied by future observations. The source showed SGR-like flaring activity in 2009 January,
observed by $INTEGRAL$ and $Swift$ (Savchenko et al. 2010).

The X-ray flux data of the candidate transient AXP $\iki$ cover a period of longer than 10 years. Tam et al.
(2006) argue that the recent flux data of $\iki$ may be from another unrelated source within the
error circles. Because of this ambiguity, we did not include this source in the present work.

The decay timescales of $\bir$ and $\uc$ are a few years (see Figures 6 and 7). The transient 
sources $\dort$ and $\alti$
seem to have been observed while still in their early decay phases ($t$ $\sim$ 150 days). 
In Figures 8 and 9, we also
present the estimated evolution of these sources in the future for different \Tcrit~values. 
We note that the decay characteristics of the model light curves depend on the
quiescent level of X-ray luminosity or accretion rate. Any corrections in distance measurements may
require a revision of some model parameters.

\section{Results and Discussion}

Our model calculations show that the idea proposed by Ertan \& Erkut (2008) to explain the X-ray
enhancement light curve of $\bir$ can be extended to other transient AXPs as well. This idea could
be summarized as follows: there is a critical temperature, \Tcrit~$\sim$ 2000 K, that prevails in
the fallback disks of all AXPs. In the X-ray enhancement phase, the viscous instability created at
this temperature governs the X-ray luminosity starting from a certain time of the decay phase,
depending mainly on the disk properties in quiescence. The properties of the extended disk, in
particular the surface density profile, can be estimated from the X-ray luminosity in quiescence,
which is different in low-luminosity transient and high-luminosity persistent AXPs. Because of these
differences in the disk properties, the effect of the instability on the decay curve is more
pronounced and starts earlier in the transient AXPs than in persistent AXPs (see Section 2.2 and Figures
3 -- 5).

A self-consistent explanation of the observed X-ray light curves requires that the basic model
parameters obtained for the different enhancement light curves of AXPs should be similar within the
uncertainties of the fallback disks. These basic parameters, which are described in detail in Section
2.2 are the viscosity parameters in the cold and hot state of the disk (\ac, \ah), the irradiation
parameter ($C$), the critical temperature (\Tcrit) and the power index ($p$) of the initial surface 
density profile, $ \Sigma \propto r^{p}, $ of the extended disk. It is seen
in Table 1 that these parameters of our model for different AXPs are either the same or very close
to each other. With these parameters, the model X-ray light curves for four different transient AXPs
are in good agreement with observations (Figures 6 -- 9).

X-ray enhancement of persistent AXPs can also be fit well by the pure viscous evolution model (Ertan
\& Alpar 2003; Ertan et al. 2006). This is consistent with our results, since the
critical temperatures found here do not significantly affect the enhancement light curves of
persistent AXPs (Section 2).

The model that we use in the present work was first proposed by Ertan \& Alpar (2003) for the
 SGR 1900+14. The only difference in our work is that we introduce a critical temperature (\Tcrit). It 
 is the presence of this \Tcrit~that leads to viscous instability during the enhancement phase. In the 
 model, this instability does not produce an X-ray outburst, but changes the evolution of the disk 
 mass-flow rate and thereby the X-ray luminosity in the decay phase of the X-ray enhancement. In the disk, 
 at temperatures below and above \Tcrit~we use different alpha parameters (\ah~and \ac) to 
 represent different viscosity states like in the models of SXTs. Over the decay 
 phase, in the disk, the radius with $T$ = \Tcrit~(cooling front) propagates inward as we explained in 
 Sections 2.2.3 and 2.2.4. In this phase, the rapid motion of the cooling front with varying X-ray 
 luminosity is a viscous instability since it changes the viscosities along the radii it propagates. On 
 the other hand, in the model, the observed X-ray enhancement is produced by the mass density and 
 temperature gradients at the inner disk.

The effect of this propagation of the cooling front on the X-ray light curve is remarkably different in 
low-luminosity transient and high-luminosity persistent systems like SGR 1900+14 and this was discussed 
in detail with illustrative model curves in the paper. All X-ray enhancement light curves of  AXP/SGRs 
mimic the light curve produced by a pure viscous decay in the early phase of evolution, but later, they 
diverge from the pure viscous decay curve due to ongoing viscous instability in the disk (Figures 3--5).  
The same \Tcrit~exists in all fallback disks, nevertheless the effect of the cooling-front propagation in 
the decay phase is more prominent and observed earlier from the systems that have lower luminosities in 
the quiescent state.  In the present work, we also present the model curves produced by pure viscous decay 
just to show, by comparison, the effect of the instability on the luminosity evolution of the sources.

The fact that Ertan \& Alpar (2003) can fit to the enhancement light curve of SGR1900+14 with a single alpha 
parameter (\ah~$\sim$ 0.1), without using a \Tcrit, indicates that the information from the cooling front 
could not communicate to the innermost disk during the observation period of this source. That is, when \Tcrit~ 
with values obtained in our work is inserted in Ertan \& Alpar (2003), their model curve does not change. 
This shows the self
 consistency of our results with the earlier work, and could also be tested by the future observations of 
 these sources. Since the critical temperature depends on the details of the chemical composition of fallback 
 disks which is not well known, it could only be estimated from the model fits. Considering that this critical
  temperature must be an intrinsic property of the fallback disks, its value must be the same in all AXP/SGR 
  systems. In the present work, we also constrain this critical temperature, trying to find solutions that can 
  fit the light curves of all these transient sources with a similar \Tcrit~along with other similar sets of 
  intrinsic disk parameters (Table 1).

In the quiescent state, the position of the cooling front remains almost constant and is determined 
by the current X-ray irradiation flux. In quiescence, back and forth motion of the cooling front in a narrow 
radial region of the disk could create variations in the local mass-flow rate (and could still be called 
viscous instability), but those are smoothed out on the way to the Alfv\'en radius and does not cause variations 
in the X-ray luminosity.  Therefore, we expect to observe the effect of the viscous instability in the decay 
phase of the X-ray luminosity.
 
We note that, in the present work, we have also refined the model parameters obtained by Ertan \&
Erkut (2008) for $\bir$ considering the newly reported last three data points of this source
(Bernardini et al. 2009). Illustrative model curves given in Ertan \& Erkut (2008) can produce the
2 -- 10 keV absorbed data\footnote[1]{There is a misprint in the label of Figure 1 in Ertan \&
Erkut (2008). The data in their Figure 1 are absorbed flux.} for a particular \ac. Nevertheless, 
these model curves are seen to remain above the new data points by a factor of three. We notice 
that the interstellar absorption 
significantly affects the light curve of the source close to the quiescent level of the X-ray luminosity. 
Here, we have repeated the calculations using unabsorbed data including the new data points of the source
(Bernardini et al. 2009). For also the other sources, except for $\alti$, we have
used unabsorbed data (Ibrahim et al. 2004; Bernardini et al. 2009; Mereghetti et al.
2006; Israel et al. 2007; Woods et al. 2011; Rea et al. 2009). Since $\alti$ is in the early decay
phase of the X-ray outburst (Figure 9), its absorbed data can be safely used to test our model.
The model curves presented in Figures 6 -- 9 are obtained with degenerate parameters \Tcrit~= 
1750 K and $C$~$\simeq$ 1 $\times$ 10$^{-4}$. For the maximum possible value of $C$ ($\sim$ 
7 $\times$ 10$^{-4}$), indicated by our earlier work, reasonable light curves could be obtained 
by increasing \Tcrit~by a factor of $\sim$ 1.6. This constrains \Tcrit~to the $\sim$ 1700 -- 2800 K range. 
This is not a strong constraint, since this result was obtained with a particular $f$ = 
\Mdot/\Mdotin~= 1 (see Section 2.1). Reasonable model curves can also be obtained with lower $f$ 
values; nevertheless this requires modification of the model parameters. For instance, using $f$ = 
0.1, a model curve that fits well to X-ray enhancement data of $\bir$ can be obtained with \ac~= 0.039 
and 1350 K $<$ \Tcrit~$<$ 2100 K, corresponding to 1 $\times$ 10$^{-4}$ $<$ $C$ $<$ 7 $\times$ 10$^{-4}$.

In Figures 6 -- 9, we also give the estimated mass \dM~of the pile-up for each of the sources. Our results 
sensitively depend on \dM. Estimated amount of burst energy \dE~imparted to the inner disk depends on \dM, 
\rin~and \rzero. Relative positions of \rin~and \rzero~also affect the model light curves, while similar 
results could be obtained with different \rin~values, adjusting \rzero~and surface density without changing 
\dM. There is an uncertainty in estimated \dE~because of the uncertainties in \rin~and surface density profile 
of the innermost disk in quiescence.

In the case of SGR 0501+4516, the distance is rather unconstrained (see Section 3). This source seems 
to be in the early phase of evolution, at a time the information from the cooling front has not reached the 
innermost disk by the viscous processes yet. This is the phase over which the light curve mimics that of a 
pure viscous decay. That is, changing the initial surface density profile at all radii by a constant 
multiplicative factor, it is possible to obtain different model light curves that have the same functional 
forms that differ only in amplitudes. Since any change in distance will modify all data points with the 
same multiplicative factor, we can obtain a similar model fit without modifying the intrinsic disk parameters.
 Nevertheless, after the viscous instability deviates the light curve from that of  pure viscous decay, any 
 correction in distances could require a modification of the critical temperature parameter to obtain a 
 similar model fit to data, or possibly the model could fail in producing the observed light curve.

We should also note the possibilities of different burst geometries. We assume that the soft gamma-ray burst 
energy is emitted isotropically. This might not be the case, at least for some of the bursts. For instance, 
only a certain angular segment of the disk could be illuminated by the burst energy. This leads to a rather 
different post-burst surface density profile than we assume here. Even in this case, the resultant enhancement 
light curves are likely to be similar to those produced by an isotropic burst, provided that the burst creates 
a sufficient surface density gradient.
These possibilities put further uncertainties on the estimated burst energy. For instance, in some cases it is 
possible that we observe an X-ray enhancement without observing the triggering burst whose anisotropic 
emission pattern evaded us. Another possibility is that we could observe 
bursts that are not followed by an enhancement in X-rays, if the solid angle of this particular burst does not 
cross the disk.

Independent of the details of burst geometries, subsequent X-ray outburst light curves of different AXPs 
provide a good test for the fallback disk model. For given quiescent and peak X-ray luminosities in an 
enhancement phase, there is a single decay curve estimated by the disk model. To put it in other words, all 
the observed enhancement light curves of AXPs should be reproduced by a single set of main disk parameters.

In comparison of the model curves with data, there are some uncertainties that we encounter at very 
low luminosities (\Lx~$\sim$ 10$^{33}$ erg
s$^{-1}$): (1) due to very low temperatures, a significant fraction of the X-ray luminosity of the
source is expected to be emitted below the observed X-ray band that we take to represent the total
luminosity of the source. (2) Absorption effects considerably increase for the soft radiation
emitted at these low temperatures. (3) Depending on the age, the cooling luminosity of
the source could have significant contribution to the total quiescent luminosity of the source.
(4) It is possible that some small bursts could be emitted in the decay phases and affect the secular decay 
characteristics of the light curves. It is not possible to address these effects in the model. For instance, 
the data point that remains above the model light curve of SGR 1627-41 (Figure 7) might be due to such a 
small burst.
 
Within these uncertainties, our model curves are in agreement with the X-ray enhancement data of four
transient AXPs (Figures 6 -- 9). We have succeeded to obtain
reasonable model curves with almost the same basic disk parameters, given in Table 1.

%% /*********************************** %% ** Conclusion ** %%

\section{Conclusions}\label{sec:conclusion}

We have shown that the X-ray outburst (enhancement) light curves of AXPs and SGRs can be explained
by the evolution of an irradiated disk after the inner disk is pushed back to larger radii by a soft
gamma-ray burst. A viscous instability created at a critical temperature, \Tcrit, seems to be a
common property of all AXP/SGR disks. For the extreme values of X-ray irradiation efficiency
obtained from our earlier work (Ertan \& \c{C}al{\i}\c{s}kan 2006), we estimate that \Tcrit~is in 
1300 -- 2800 K range.

Characteristic differences between the enhancement light curves of transient and
persistent AXP/SGRs can naturally be accounted for by their different
pre-burst (quiescent) conditions of the disks implied by the X-ray luminosity of the sources in
quiescence. X-ray outburst light curve of a persistent AXP/SGR could not be distinguished from a
light curve produced by a pure viscous (without any instability) evolution of the disk for a few
years. For a transient source, the outburst light curve could diverge from the pure viscous
evolution within months after the onset of the outburst (Figures 6 -- 9).
These results are consistent with our earlier work on the X-ray outburst light curves of persistent
AXP/SGRs (see, e.g., Ertan et al 2006) which were explained by pure viscous evolution of the disk.

Basic properties of the fallback disks are likely to be similar in the fallback disks of all
AXP/SGRs. Through a large number of
simulations, we have obtained a single set of these basic parameters (Table 1) that can produce
reasonable model fits to the enhancement light curves of four transient AXP/SGRs
(Figures 6 -- 9).

The predictions of our model could be tested by future observations of AXPs and SGRs in the X-ray 
enhancement phases.

%% /********** %% ** Acknowledgments ** %%

\acknowledgements

We acknowledge research support from T\"{U}B{\.I}TAK (The Scientific and Technical Research Council
of Turkey) through grant 110T243 and from the Sabanc\i\ University Astrophysics and Space Forum.
This work has been supported by the Marie Curie EC FPG Marie Curie Transfer of Knowledge Project
ASTRONS, MKTD-CT-2006-042722. We thank Ali Alpar, Hakan Erkut and Yavuz Ek\c{s}i for useful
discussions and comments on the manuscript. We thank the anonymous referee for his/her
useful comments.

\clearpage 
\begin{figure}[t]
\centerline{\includegraphics[width=0.75\textwidth,angle=270]{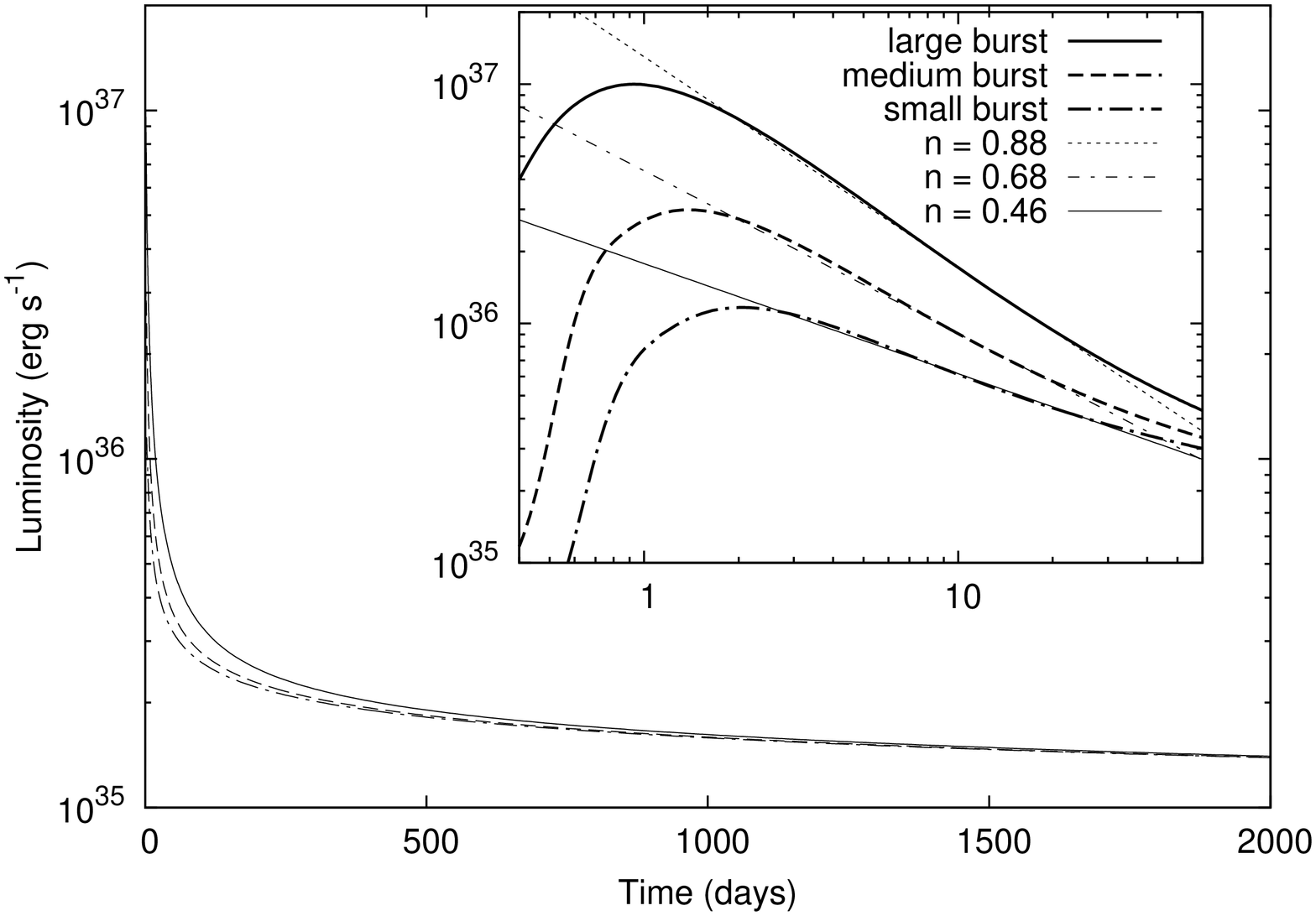}}
\caption{\label{fig:psds} 
Model light curves produced by pure viscous evolution of disks
for three different \dE~values. The short-term light curve, in the inset, shows peak luminosities of
10$^{37}$ erg s$^{-1}$, 3 $\times$ 10$^{36}$ erg s$^{-1}$, and 10$^{36}$ erg s$^{-1}$, which all
decay to the same quiescent luminosity in the long term. The \dM~ values are 4.9
$\times$ 10$^{22}$ g, 2.1 $\times$ 10$^{21}$ g, and 8.9 $\times$ 10$^{20}$ g and the estimated \dE~values
are 9.2 $\times$ 10$^{39}$ erg, 3.8 $\times$ 10$^{39}$ erg, and 1.7 $\times$ 10$^{39}$ erg
respectively. The model light curves can be fitted with power laws in the early decay phase ($\sim$
a few weeks). The values of power indices ($n$) are given in the inset. } 
\end{figure}

\clearpage 
\begin{figure}[t]
\centerline{\includegraphics[width=0.75\textwidth,angle=270]{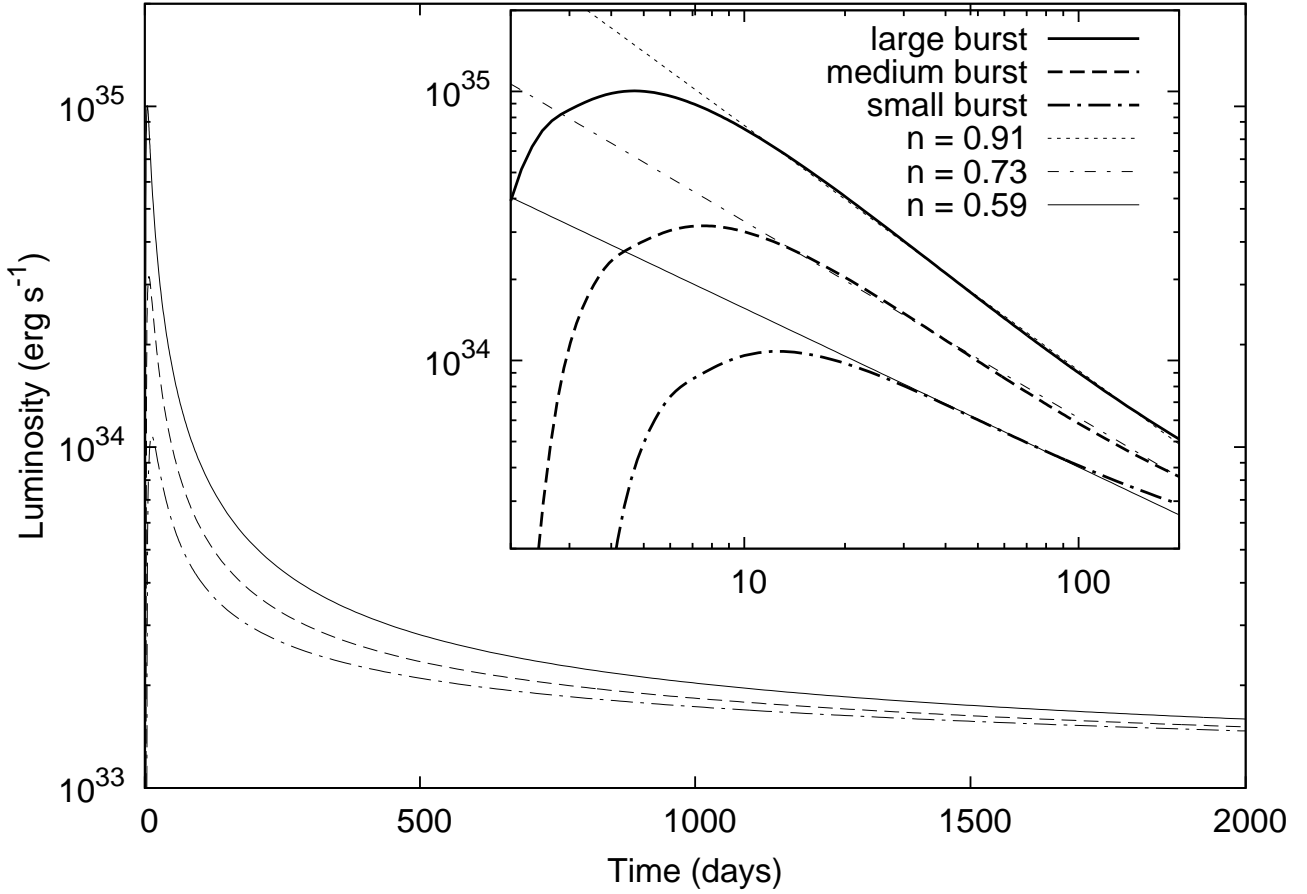}}
\caption{\label{fig:psds} 
Short-term and long-term model light curves of a typical transient source
for three different \dE~values. For these models, \dM~ values are 2.4 $\times$ 10$^{21}$ g, 1.2
$\times$ 10$^{21}$ g, and 6.4 $\times$ 10$^{20}$ g and estimated \dE~values are 4.5 $\times$
10$^{38}$ erg, 2.3 $\times$ 10$^{38}$ erg, and 1.2 $\times$ 10$^{38}$ erg respectively. The decay
phases of the light curves can be fitted with power laws for the first $\sim$ 100 days (inset). The
values of power indices ($n$) are given in the inset. } 
\end{figure}

\clearpage 
\begin{figure}[t]
\centerline{\includegraphics[width=0.95\textwidth,angle=270]{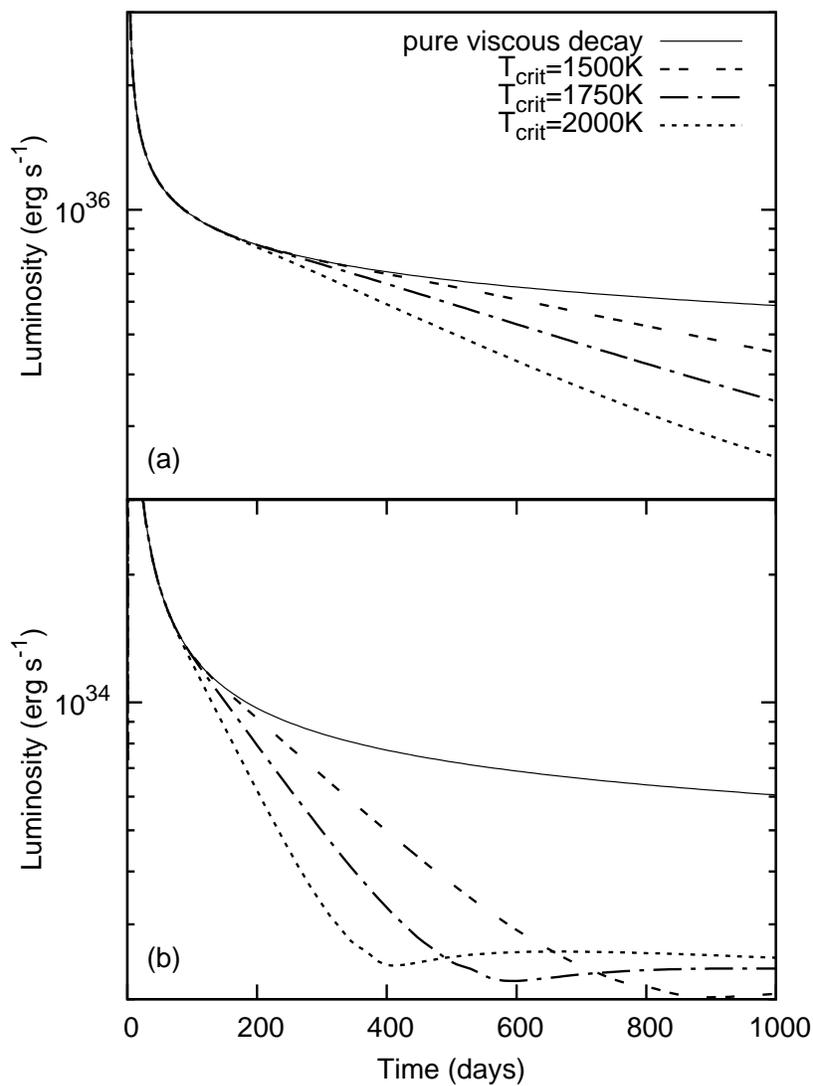}} 
\caption{\label{fig:psds}
Model light curves for persistent (a) and transient (b) sources, for \Tcrit = 1500 K, 1750 K, and 2000
K. Solid curves illustrate the pure viscous decay with the same initial conditions for comparison.
For a given \Tcrit, comparing (a) and (b), the difference between the model curves of the transient and persistent sources is clearly seen.
} 
\end{figure}

\clearpage 
\begin{figure}[t]
\centerline{\includegraphics[width=0.95\textwidth,angle=270]{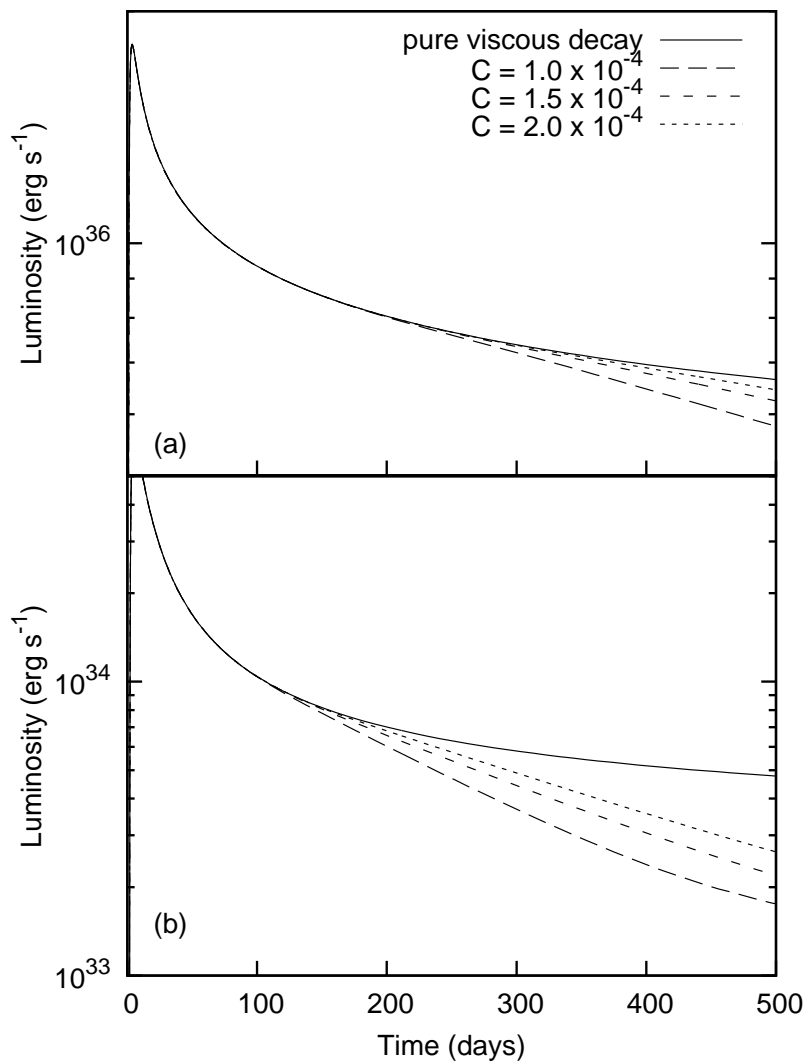}} 
\caption{\label{fig:psds}
Top panel shows the model light curves of a persistent source for different $C$ values, and the
bottom panel shows the light curves for a transient source, for the dame parameter values. 
The model curves representing 
pure viscous decay are given with solid lines and \Tcrit~ = 1750 K for the other models. It is seen that 
the evolution of persistent sources diverges from the pure viscous decay curve much later than transients.
} 
\end{figure}

\clearpage 
\begin{figure}[t]
\centerline{\includegraphics[width=0.95\textwidth,angle=270]{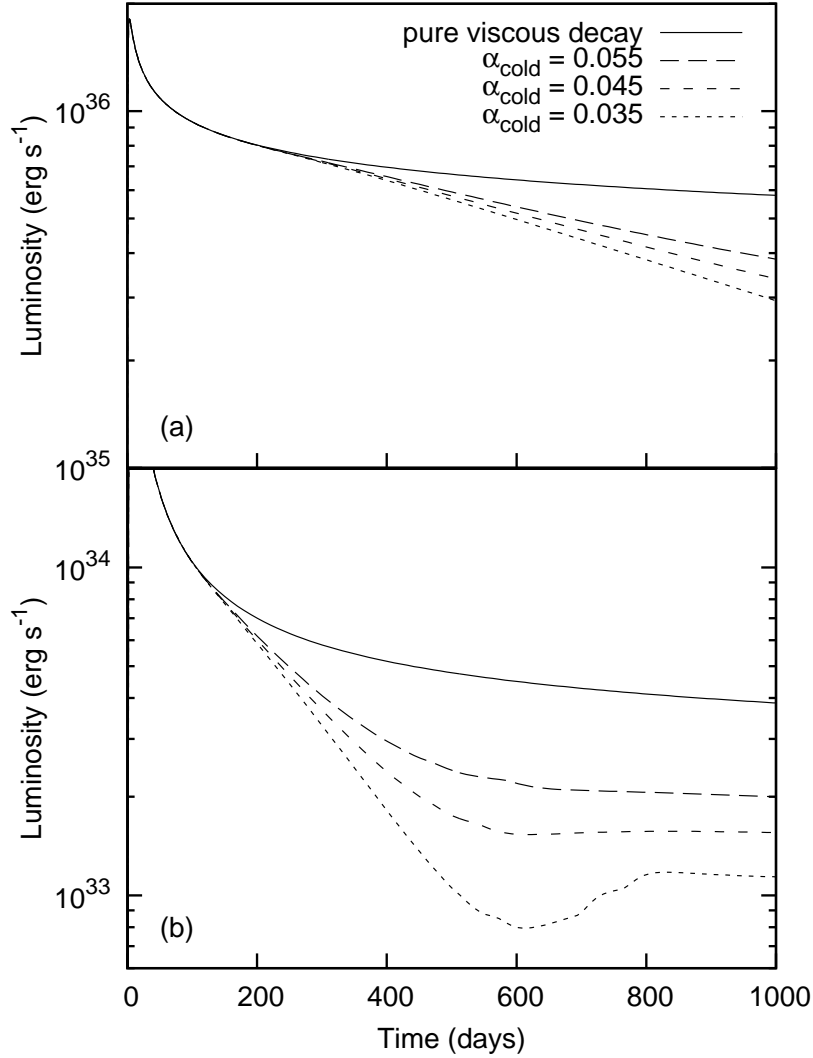}}
\caption{\label{fig:psds} Top panel shows the light curve of a persistent source for different
\ac~ values, and the bottom panel shows the light curves for a transient source, for the dame parameter 
values. The pure
viscous decay curves are also presented (solid lines) for comparison. For the other models, we take
 \Tcrit~= 1750 K and $C$ = 1 $\times$ 10$^{-4}$. 
} 
\end{figure}

\clearpage 
\begin{figure} \includegraphics[height=.8\textheight,angle=270]{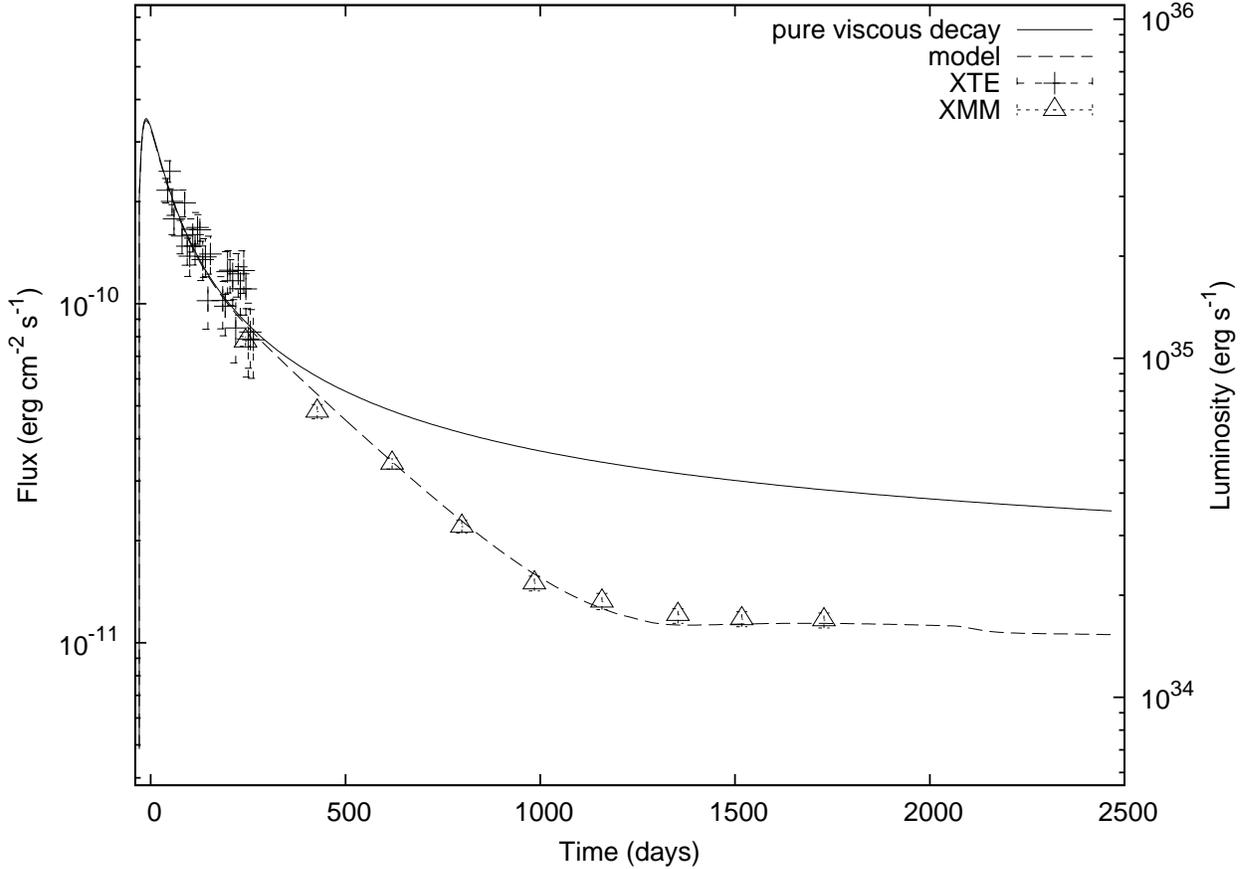}
\caption{0.1 -- 10 keV unabsorbed X-ray flux and luminosity data of XTE J1810--197. The absorbed
0.6 -- 10 keV $XMM$ data (Bernardini et al. 2009) were converted to unabsorbed 0.1 -- 10 keV flux using the 3BB model described in
their paper. The XTE data were given in counts s$^{-1}$ cpu$^{-1}$ with a conversion factor for 2 -- 10 keV
absorbed flux (Ibrahim et al. 2004), with no spectral fits. The XTE data were rescaled by
a factor of 2.3 to match the first $XMM$ data, taken in 2003 September (see the text for details). 
The luminosity is
calculated assuming $d$ = 3.5 kpc. For this model,  \dM~$\sim$ 1 $\times$ 10$^{23}$ g and the estimated
\dE~$\sim$ 1 $\times$ 10$^{40}$ erg.} 
\end{figure}

\clearpage 
\begin{figure} \includegraphics[height=.8\textheight,angle=270]{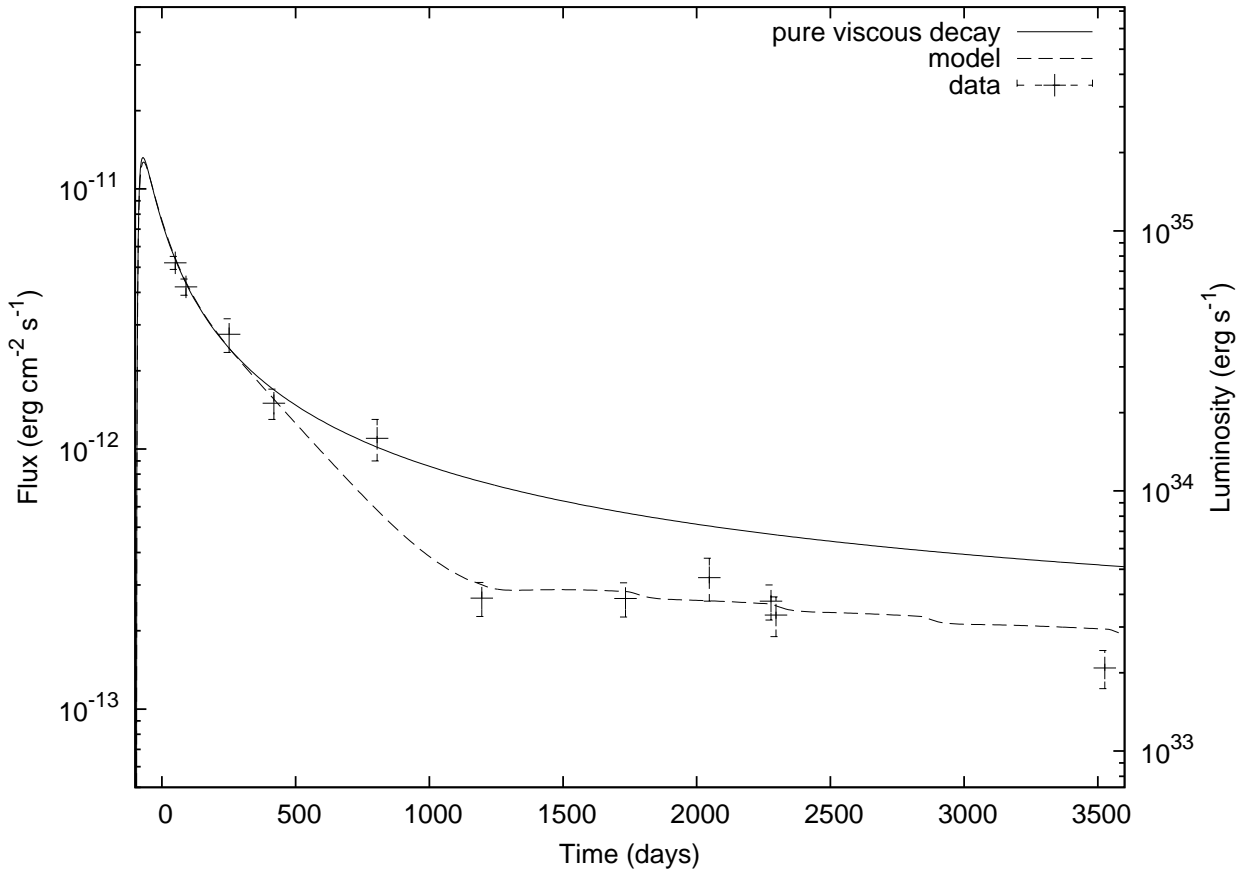}
\caption{Unabsorbed 2 -- 10 keV flux and luminosity data of SGR 1627--41 (Mereghetti et al.
2006). The parameters of the model curve, given by the dashed line, are listed in Table 1. The solid curve 
represents pure viscous decay with the same initial conditions. 
These model curves are obtained with 
\dM~$\sim$ 4 $\times$ 10$^{22}$ g, which gives \dE~$\sim$ 4 $\times$ 10$^{39}$ erg with the chosen 
\rin~(see the text for details). The luminosity is calculated assuming $d$ = 11 kpc. 
}
\end{figure}

\clearpage 
\begin{figure} \includegraphics[height=.8\textheight,angle=270]{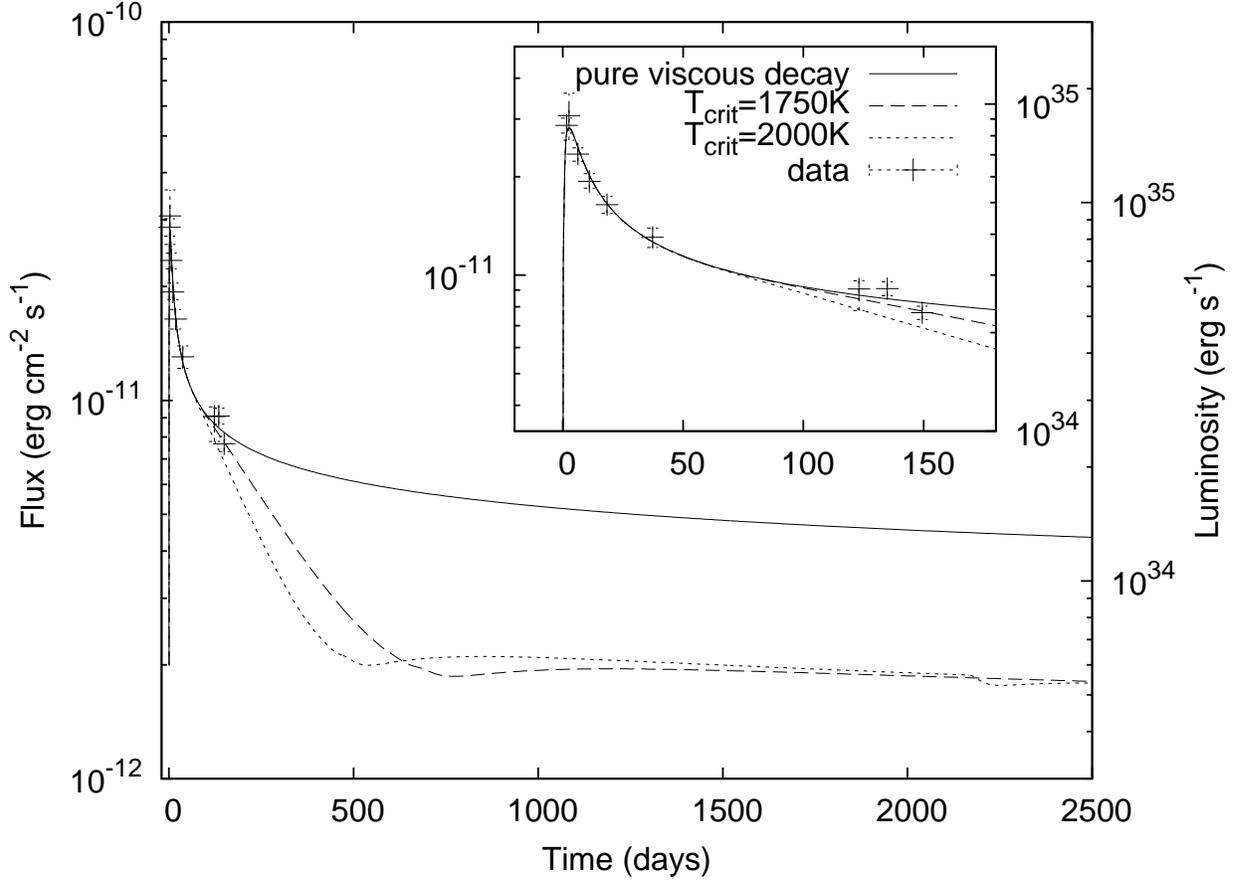}
\caption{Unabsorbed 2 -- 10 keV flux data of $\dort$ (Israel et al. 2007; Woods et al. 2011).
The long-term model light curves are obtained with \Tcrit = 1750 K, and \Tcrit = 2000 K. For these models,
\dM~$\sim$ 2 $\times$ 10$^{21}$ g and the estimated \dE~$\sim$ 2 $\times$ 10$^{38}$ erg.
The luminosity is calculated assuming $d$ = 5 kpc. 
} 
\end{figure}

\clearpage 
\begin{figure} \includegraphics[height=.8\textheight,angle=270]{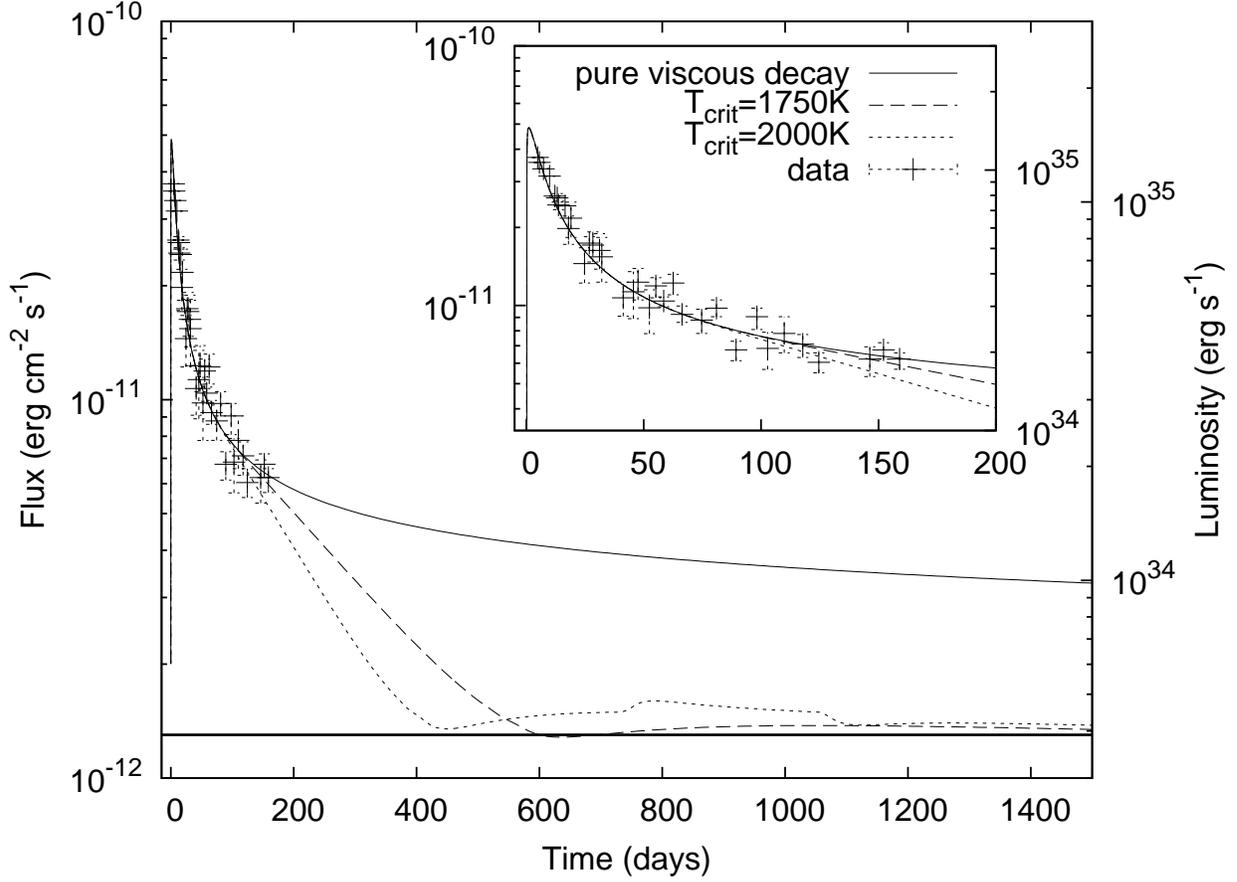}
\caption{Absorbed 1--10 keV flux and luminosity data of SGR 0501+4516 (Rea et al. 2009). 
The model light curves are obtained with \Tcrit =
1750 K, and \Tcrit = 2000 K. The horizontal line shows the estimated quiescent flux of the source 
(1.3 $\times$
10$^{-12}$ erg cm$^{-2}$ s$^{-1}$), obtained from $ROSAT$ observations in 1992, extrapolated to the 1--10 keV band assuming a blackbody emission (Rea et al. 2009).
The luminosity is calculated assuming $d$ = 5 kpc. 
These model curves are obtained with \dM~$\sim$ 3 $\times$ 10$^{21}$ g. We estimate \dE~$\sim$ 
2 $\times$ 10$^{38}$ erg. It is seen that the source is about to diverge from the pure viscous 
decay curve (solid curve).
}
\end{figure}

\clearpage 
\begin{table*}[t] \caption{\label{table:params}The parameters for the model curves
presented in Figures 6--9. Note that the parameters \ah, \ac, $p$
and \Tcrit~ are expected to be similar for all AXPs and SGRs. Irradiation efficiency, $C$, which
could change with accretion rate is also likely to be similar for the sources in the same accretion
regimes. In quiescence, \Szero~scales with accretion rate. The parameters \dr, \rzero, and \Smax~could
vary from source to source, depending on the burst energy and geometry. 
The values of inner disk radius \rin~are close to the Alfv\'en radii of the sources with \Bzero~$\simeq$ 
10$^{12}$ G. We set \rout~= 10$^{13}$ cm and $f$ = \Mdot/\Mdotin~= 1 for all our models.
}

\begin{minipage}{\linewidth} 
\begin{center} 
\begin{tabular}{c|c|c|c|c} \hline \hline 
Parameter & XTE
J1810-197 & SGR 1627-41 & CXO J164710.2-455216 & SGR 0501+4516\\ \hline 
\rin~(cm) & 2 $\times$ 10$^{9}$ & 2 $\times$ 10$^{9}$ & 1.8 $\times$ 10$^{9}$ & 3 $\times$ 10$^{9}$\\ 
\rzero~(cm) & 7 $\times$ 10$^{9}$ & 2.3 $\times$ 10$^{10}$ & 5 $\times$ 10$^{9}$ & 6 $\times$ 10$^{9}$ \\ 
\dr (cm) & 9 $\times$ 10$^{9}$ & 6 $\times$ 10$^{8}$ & 6 $\times$ 10$^{8}$ & 1.4 $\times$ 10$^{9}$ \\
\Smax (g cm$^{-2}$) & 20 & 60 & 13 & 10\\ 
\Szero (g cm$^{-2}$) & 10 & 1 & 7.6 & 5\\ 
\ah & 0.1 & 0.1 & 0.1 & 0.1\\
\ac & 0.045 & 0.045 & 0.045 & 0.045\\ 
\Tcrit~(K) & 1750 & 1750 & 1750 & 1750\\ 
$C$ & 1 $\times$ 10$^{-4}$ & 1.6 $\times$ 10$^{-4}$ & 1 $\times$ 10$^{-4}$ & 1 $\times$ 10$^{-4}$ \\ 
$p$ & 0.75 & 0.75 & 0.75 & 0.75\\ 
\hline \hline 
\end{tabular}\\

\end{center} \end{minipage} \end{table*}

\end{document}